\documentclass[prb,onecolumn,nofootinbib,citeautoscript,10pt,longbibliography,notitlepage]{revtex4-2}

\usepackage{graphicx}
\usepackage{dcolumn}
\usepackage{color}
\usepackage{amssymb,amsmath}
\usepackage{tabularx,graphicx}
\usepackage{epstopdf}
\usepackage{latexsym}
\usepackage{colortbl}
\usepackage{psfrag}
\usepackage{bbm,bm,array,physics,xparse}
\usepackage{titlesec}
\usepackage{dsfont}
\usepackage[tight]{subfigure}
\usepackage[pdftex,colorlinks=true,linkcolor=darkblue,
citecolor=blue,urlcolor=darkred]{hyperref}

\usepackage[papersize={8.5in,11in}]{geometry}
\definecolor{darkblue}{rgb}{0.,0.,0.4}
\definecolor{darkred}{rgb}{0.5,0.,0.}
\definecolor{BlueViolet}{RGB}{138,43,226}
\definecolor{SkyBlue}{RGB}{30,144,255}
\definecolor{DarkGreen}{RGB}{0,100,0}

\def \nn{\nonumber \\}
\def\*#1{\mathbf{#1}} 
\def\t#1{\text{#1}} 
\def\mc#1{\mathcal{#1}} 
\def\lrb{\left(} 
\def\rrb{\right)} 
\def\lsb{\left[} 
\def\rsb{\right]} 
\def\lp{l^{\prime}}
\def\np{n^{\prime}}
\def\sp{s^{\prime}}
\def\ev{\text{eV}}

\def\adag{a^{\dagger}}
\def\nmax{n_{\text{max}}}
\newcommand\pmtx[1]{\begin{pmatrix}#1\end{pmatrix}} 
\newcommand\perm[2]{^{#1}P_{#2}}
\newcommand\dperm[2]{^{#1}\Gamma_{#2}}
\newcommand\kr[2]{\delta_{#1,#2}}

\begin{document}

\title{Magneto-optical conductivity in the type-I and type-II phases of Weyl/multi-Weyl semimetals}

\author{Shivam Yadav}

\affiliation{Institute of Nuclear Physics, Polish Academy of Sciences, 31-342 Krak\'{o}w, Poland}

\author{Sajid Sekh}
\affiliation{Institute of Nuclear Physics, Polish Academy of Sciences, 31-342 Krak\'{o}w, Poland}

\author{Ipsita Mandal}

\affiliation{Institute of Nuclear Physics, Polish Academy of Sciences, 31-342 Krak\'{o}w, Poland}

\begin{abstract}
Magneto-optical conductivity is a very widely studied transport coefficient, useful to understand and characterize the behaviour of materials under magnetic fields. Using the Kubo formula, we compute the components of the conductivity tensor $\sigma_{\mu \nu}$ transverse to a uniform magnetic field $\mathbf B$, for a single node of a multi-Weyl semimetal (with monopole charge $J$ equal to two or three). 
We also include the results for a Weyl semimetal (with $J=1$), and identify peaks in the conductivity profile which were not reported in earlier studies. In our analysis, we explore how the tilting of the Weyl/multi-Weyl cone affects $\sigma_{\mu \nu}$, focussing on both type-I and type-II phases. All these systems have a linear-in-momentum dispersion along the tilting axis, which is chosen to align with $\mathbf B$.
In the type-II phases, open Fermi pockets appear as artifacts of the low-energy effective continuum models, which ignore higher-order momentum terms of the actual bandstructures. Hence, we supplement the linear power term with a cubic term, which closes the Fermi pockets, thus eliminating any need for an ad hoc cutoff for the momentum integrals. Our results reveal that the absorptive parts of $\sigma_{\mu \nu}$ display multiple peaks as functions of the frequency, whose locations are determined by an energy scale $\sim |\mathbf B|^{J/2} $.
\end{abstract}
\maketitle

\tableofcontents

\section{Introduction}
\label{sec:intro}

Three-dimensional (3d) gapless semimetallic phases, which also exhibit nontrivial topology in their bandstructures, encompass systems with multiple band-crossings in the Brillouin zone (BZ). In the simplest case, the low-energy dispersion of such a system contains two bands crossing each other linearly at a nodal point. These are commonly known as ``Weyl semimetals'' (WSMs) \cite{hosur2013recent,vafek2014,wan2011topological,burkov2011,armitage2018,lv2015experimental}, since the energy bands around such a node show a linear momentum dependence, reminiscent of the relativistic Weyl fermions in high-energy physics. A Weyl node acts as a source or sink of Berry curvature, and gives rise to quantized flux in integer multiples of $2\pi$. The quantization is characterized by the topological-invariant called the ``Chern number'', which takes only integer values.

In addition to the most familiar example of WSMs that exhibit isotropic linear dispersion, multi-WSMs~ \cite{liu2017predicted,lundgren2014thermo,fang2012multi} are their cousins having a mix of linear and higher-order dispersions, depending on the directions. For example, a double(triple)-WSM harbours a linear dispersion along one direction, but a quadratic(cubic) dispersion in the plane perpendicular to it. These are all two-band 3d semimetals hosting Chern numbers $\pm J$, where
$J$ takes the values $1$, $2$, and $3$, for WSM, double-WSM, and triple-WSM, respectively. The  Nielsen-Ninomiya theorem imposes the condition that the nodal points always appear in pairs~\cite{nielsen-ninomiya}, featuring equal and opposite Chern numbers, which ensures that the net topological charge vanishes for the full BZ. 

Let us choose the direction of the linear dispersion to be along the momentum component $k_z$ for each of these nodal-point semimetals. In generic situations, imperfections are ubiquitous, and there can be a tilt in the energy dispersion with respect to this momentum component. If the tilt is small enough such that the chemical potential cutting the nodal point gives a Fermi point rather than a Fermi surface (FS), it is referred to as the type-I phase. However, if the tilt is so large such that when the Fermi level cuts it, the nodal-point appears at the contact point between electron-like and hole-like pockets, we call it a type-II phase \cite{soluyanov2015type}. The over-tilting leads to the formation of Fermi pockets.

\begin{figure}[]
	\centering
	\includegraphics[width=0.8\linewidth]{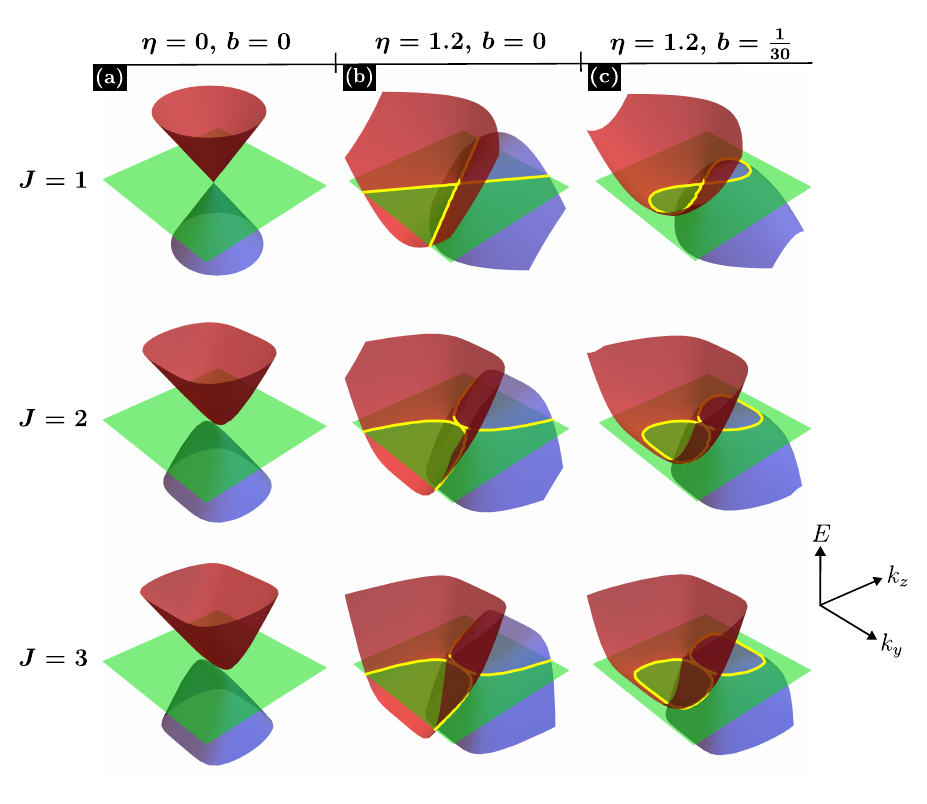}
	\caption{Schematic bandstructures of WSM ($J=1$), double-WSM ($J=2$), and triple-WSM ($J=3$) are shown for the Hamiltoinian in Eq.~\eqref{eq:ham}. The three vertical panels represent three different scenarios for the tilt and cubic correction parameters: (a) untilted cone without cubic correction ($\eta = b=0$);
(b) over-tilted cone without cubic correction ($\eta=1.2$ and $ b=0$);
and (c) over-tilted cone with cubic correction ($\eta=1.2$ and $ b= 1/30$).
Over-tilting gives rise to Fermi pockets, which are open in the absence of a higher-order momentum correction, as seen in panel (b). A finite cubic correction in panel (c) closes these Fermi pockets, which is the realistic representation of materials found in nature.
\label{fig:disp}
}
\end{figure}

The magneto-optical response has proved to be a versatile tool, which can be used to characterize various materials like graphene, WSMs, double-WSMs, and nodal-line semimetals \cite{gusynin2006magneto,sharapov-graphene,
fuseya2011spin,tabert2013valley,li2013magneto,carbotte13,magneto-double-weyl,marcus-emil,zhao_nodal, 
nodal-line,nodal-line-type2}. In this paper, we supplement the earlier investigations involving WSMs/double-WSMs by adding suitable corrections to the dispersion along the $z$-direction, which are higher-order in momentum, following the analysis in Ref.~\cite{marcus-emil}. Moreover, we include the triple-WSM case.
Inclusion of higher-order corrections, which are naturally expected to be present in a realistic scenario, 
has the advantage that they close the Fermi pockets in the type-II regime. Indeed, in reality, we do have closed Fermi pockets, rather than the unphysical open (or unbounded) ones, as the latter are the artifacts of the effective continuum models valid only in the low-energy regimes. Hence, this procedure eliminates the need for unphysical momentum (or energy) cutoffs while performing the momentum integrals, appearing in the conductivity expressions. The behaviour of the dispersion under the interplay of the tilt parameter and a cubic correction is shown schematically in Fig.~\ref{fig:disp}.

The paper is organized as follows. In Sec.~\ref{sec:model}, we show the Hamiltonian capturing the low-energy continuum model for a Weyl/multi-WSM, and discuss the formation of Landau levels under the influence of an external magnetic field. In Sec.~\ref{sec:cond}, we elucidate the expressions for the magneto-optical conductivity for these semimetals. This is followed by an analytical calculation that discusses the location of the conductivity peaks due to Landau levels. We show our numerical results in Sec.~\ref{sec:result}. In particular, Sec.~\ref{sec:typeI} and Sec.~\ref{sec:typeII} contain the results for the type-I and type-II phases, respectively. We also illustrate the features of the conductivity for polarized light in Sec.~\ref{sec:pol}. Finally, we conclude with a summary and some discussions in Sec.~\ref{sec:conclusion}.

\section{Landau levels: Weyl and multi-Weyl semimetals under magnetic field}
\label{sec:model}

The low-energy effective Hamiltonian for a single WSM/multi-WSM node, with chirality $\chi$,
is given by \cite{liu2017predicted,fang2012multi}
\begin{align}
	\mc{H}(\*k) = h(k_z) \, \mathbb{I}_2 
+ \chi \,\alpha_J \, k_{\perp}^J \left[  \cos(J\phi_\*k) \, \sigma_x + 
	\sin(J\phi_\*k) \, \sigma_y \right ] 
+ \chi \,	\, g(k_z) \, \sigma_z \, ,
	\label{eq:ham}
\end{align}
where $\{\sigma_x, \sigma_y, \sigma_z\}$ are Pauli matrices, $\mathbb{I}_2$ refers to the $2\times 2$ identity matrix, $h(k_z)$ and $g(k_z)$ are continuously differentiable functions of $k_z$, and $J\in \{1,2,3\}$ describes the topological charge of the nodal-point. Here, we will just focus on the positive chirality node, corresponding to $\chi = 1$. Given that $\*k \in \{k_x,k_y,k_z\}$ represents the wavevector in three spatial dimensions, we have defined $k_{\perp}= \sqrt{ k_x^2+k_y^2}$ and $\phi_\*k = \arctan(k_y/k_x)$ for compactness of notations. Additionally,
$\alpha_J = (v_{\perp}/k_0^{J-1})$ is a material-dependent constant, with $v_{\perp}$ denoting the Fermi velocity in the $xy$-plane. For simplicity, we resort to the natural units $\hbar=c=k_B=1$, and set $v_{\perp}=0.5$ in all the expressions that follow. 

The tilting of the cone along the $z$-direction is tuned by the term $h(k_z)$, which we choose to be equal to $ \eta \, v_z \, k_z$, where $|\eta|<1$ and $|\eta|>1$ refer to the type-I and type-II phases, respectively.
In our explicit calculations, we set $g(k_z)= v_z \, (k_z + b \, k_z^3)$, where $v_z$ is the Fermi velocity along $z$-direction, and assume it to be equal to $v_{\perp}$ for the sake of simplicity. The magnitude of $b$ controls the strength of the cubic corrections. The purpose of considering a nonzero cubic correction is to avoid any unphysical cutoff in the momentum integrals~\cite{marcus-emil}, as explained in the introduction.

In order to compute the characteristic magneto-optical conductivity, the system is subjected to a uniform external magnetic field $\mathbf{B}=(0,0,B)$ (choosing $B>0$), which is aligned along the $z$-direction. The presence of this magnetic field introduces a vector potential $\mathbf{A}$, such that $\mathbf B = \boldsymbol{\nabla} \times \mathbf A$. This is incorporated into the system by replacing $\mathbf{k} \rightarrow \mathbf{\Pi} = \mathbf{k} + e \, \mathbf{A}$. For our purpose, we choose the Landau gauge, which implies $\mathbf A = (0, B \, x, 0)$. Additionally, we introduce the creation $a^{\dagger} = \frac{l_B}{\sqrt{2}}(\Pi_x + i \, \Pi_y)$ and annihilation $a = \frac{l_B}{\sqrt{2}}(\Pi_x - i \, \Pi_y)$ operators, obeying the commutation relation $\left[ a,a^{\dagger} \right] =1$, where $l_B =1/\sqrt{e\, B}$ is the magnetic length scale. Using these ingredients, the Hamiltonian takes the form
\begin{align}
\mc{H}_B(\*k) 
=
\begin{pmatrix}  
	 g( k_z ) + h( k_z ) &  \lambda_J \, a^J \\ 
		\lambda_J \, (a^\dagger)^J  & -  g( k_z) +  h( k_z )
	\end{pmatrix}		\, ,
\quad \lambda_J= \left (\sqrt{2}/l_B \right )^J \, \alpha_J\,,		
	\label{eq:ham2}
\end{align} 
in the presence of the magnetic field. The energy eigenvalues are given by (see Appendix~\ref{appx:eigensystem} for more details)
\begin{align}
	E_{n,s} (k_z) = \begin{cases}
h(k_z)+g(k_z) & \text{ for } n = 0, 1, \cdots , J-1 \\
h(k_z) + s\, \Delta_n(k_z) & \text{ for } n \geq J 
	\end{cases}	 \,,
\quad \Delta_n(k_z) = \sqrt{g^2 (k_z) + \lambda_J^2 \, \perm{n}{J}}\,.	
\label{eqevs}	
\end{align}
Here $\perm{n}{J} = n!/(n-J)!$ , $n$ is the band-index, and $s=\pm $ labels whether we are considering a conduction ($+$) or valence ($-$) band. Without any loss of generality, we can express the corresponding eigenstates as
\begin{align}
\label{eqestates}
& \ket{\Psi_{n,s}}=
	\begin{cases}
		\pmtx{0 &	 1}^T  & \text{ for } n = 0, 1, \cdots , J-1 \\
		\pmtx{s \, \mc{U}_{n,s} & \mc{V}_{n,s}}^T   & \text{ for } n \geq J	
	\end{cases} \, ,
\quad \mc{U}_{n,s}=u_{n,s} \, \Phi_{n-J} \,,\quad \mc{V}_{n,s}=v_{n,s} \, \Phi_{n}\,,
\nn & u_{n,s}(k_z) = \sqrt{\frac{1}{2} \, \lsb 1+ \, \frac{g(k_z)}{E_{n,s}(k_z)-h(k_z)} \rsb} \, , \quad
	v_{n,s}(k_z) = \sqrt{\frac{1}{2} \, \lsb 1- \, \frac{g(k_z)}{E_{n,s}(k_z)-h(k_z)} \rsb} \, .	
\end{align}
The symbol $\Phi_n$ represents the $n^{\t{th}}$ eigenstate of a simple harmonic oscillator  (see the Appendix~\ref{appx:eigensystem} for details).

\section{Magneto-optical conductivity tensor}
\label{sec:cond}

Within the linear response theory, the expression for the magneto-optical conductivity tensor
\begin{align}
\label{eqkubo}
\sigma_{\mu\nu}(\omega) = -\frac{i} {2\,\pi \,l_B^2} \, \sum_{l,\lp} \, 
	\int \frac{dk_z}{2\pi} \, 
	\lsb \frac{f_{l}(k_z) - f_{\lp}(k_z)}{E_l(k_z) - E_{\lp}(k_z)} \rsb \,
	\frac{\mc{M}^{l, \lp}_{\mu\nu}(k_z)}
	{\omega - E_{l}(k_z) + E_{\lp}(k_z) + i \, \epsilon} \,,
\text{ with } \mu, \nu \in \lbrace x, y, z \rbrace	\,,
\end{align}
is given by the Kubo formula~\cite{carbotte13}, where $\omega $ denotes the frequency of the absorbed/emitted photon. With $l  \equiv \{n, s\}$ denoting the sum index, the elements of the $\mc{M}$-matrix are given by $\mc{M}^{l,\lp}_{\mu\nu} = \expval{\Psi_{l}|J_{\mu}|\Psi_{\lp}} \expval{\Psi_{l}|J_{\nu}|\Psi_{\lp}}$, where $J_{\mu} = e \, (\partial \mc{H}_B/\partial \Pi^{\mu})$ is the current density operator, and $\epsilon$ denotes the impurity scattering rate \footnote{For the sake of simplicity, we assume that all Landau levels have the same value of $\epsilon$.}. The Fermi-Dirac distribution is given by $f_l(k_z) = 1/\lsb 1+ 
e^{\beta\left \lbrace  E_l(k_z)-\mu \right \rbrace } \rsb$, where $\mu$ is the chemical potential, and $\beta=1/T$ is the inverse temperature. Here, we will calculate the diagonal and off-diagonal parts of the transverse components (i.e., perpendicular to the $z$-direction) of the conductivity tensor. Due to the unbroken rotational symmetry in the $xy$-plane, $\sigma_{xx} = \sigma_{yy}$ and $\sigma_{xy} = \sigma_{yx}$. Hence, we will derive the expressions for $\sigma_{xx}$ and $\sigma_{xy}$ only.

Using the current density-expressions
\begin{align}
	J_x = \frac{J \, \lambda_J \,  l_B \, e}{\sqrt{2}} \, 
	\pmtx{0 && a^{J-1} \\
		(a^{\dagger})^{J-1} && 0}  \text{ and }
	J_y = \frac{J \, \lambda_J \,  l_B \, e}{\sqrt{2}} \, 
	\pmtx{0 && -i \, a^{J-1} \\
		i \, (a^{\dagger})^{J-1} && 0} \, ,
	\label{eq:j-matrix}
\end{align}
along the  $x$- and $y$-directions, respectively,
we find that
\begin{align}
	&\mc{M}^{l,\lp}_{xx} = \frac{e^2 \, \lambda_J^2 \, l_B^2 \,  J^2}{2} \, 
	\lrb 
	u_{n,s}^2 \, v_{\np,\sp}^2 \, \perm{\np}{J-1} \, \kr{n}{\np+1} +
	u_{\np,\sp}^2 \, v_{n,s}^2 \, \perm{\np}{J-1} \, \kr{n}{\np-1}
	\rrb \, , 
	\nn
	&\mc{M}^{l,\lp}_{xy} = \frac{i \, e^2 \, \lambda_J^2 \, l_B^2 \,  J^2}{2} \, 
	\lrb 
	-u_{n,s}^2 \, v_{\np,\sp}^2 \, \perm{\np}{J-1} \, \kr{n}{\np+1} +
	u_{\np,\sp}^2 \, v_{n,p}^2 \, \perm{\np}{J-1} \, \kr{n}{\np-1}
	\rrb \, .
\label{eq:m-matrix}
\end{align}
Each Kronecker delta function in the matrix elements gives the selection rule for the transition between the Landau levels, enforcing the condition that $ n' = n\pm 1$ contributes to the nonzero terms.
Note that we have suppressed the $k_z$-dependence here, in order to avoid cluttering of notations. We will follow this convention in the rest of the paper. 

\begin{figure}[]
\centering
\includegraphics[width=0.9\columnwidth]{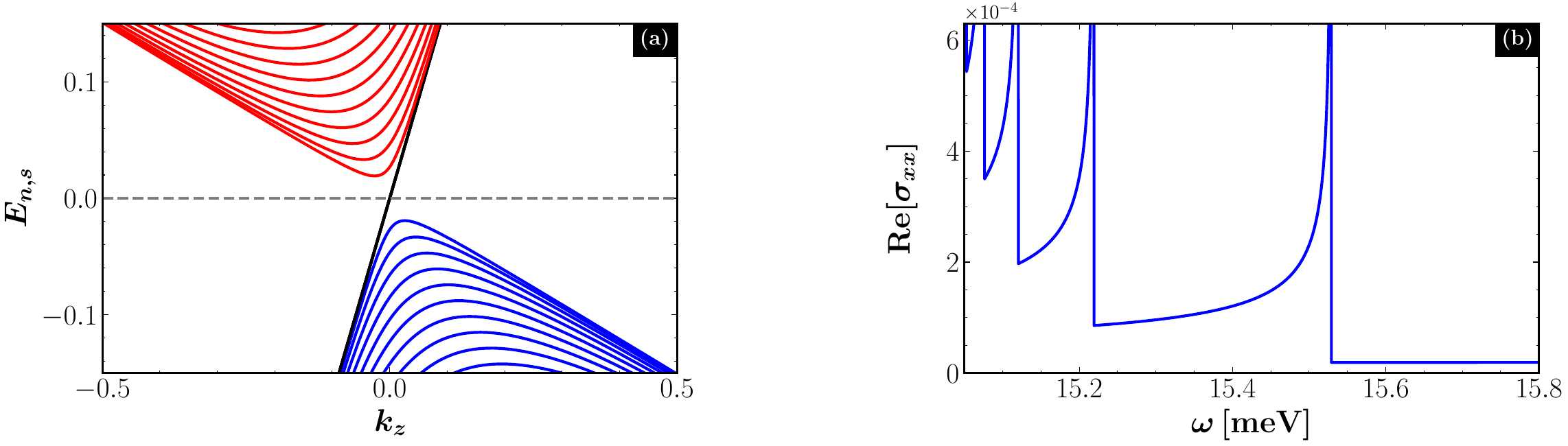}
\caption{Characteristics for a double-WSM ($J=2$) with $\alpha_2 =0.012 \, \ev^{-1}$, in the type-I phase ($\eta=0.7$) without the cubic correction ($b=0$) [cf. Eq.~\eqref{eqevs}], when subjected to a magnetic field of magntitude $B=6.25 \, \ev^2$ at $\beta=100 \, \ev^{-1}$ (resulting in $\lambda_2= 15\,\mathrm{meV}$).
Subfigure (a) shows the dispersion (in eV) of the first few Landau levels, as a function of the momentum $k_z$ (in eV). The grey dashed line indicates the Fermi level, which is set to $\mu=0$. Subfigure (b) is a plot of the real part of $\sigma_{xx}$ [in units of $\ev^2$] as a function of frequency $\omega$, focussing on the region $\omega \sim \lambda_2$, in order to capture the contributions from the conduction-to-conduction band transitions. Here, $\mu$ is set to zero.
\label{fig:cond_cc}}
\end{figure}

As the final step, we substitute Eq.~\eqref{eq:m-matrix} in the conductivity tensor to compute the transverse components in the clean limit (captured by $\epsilon \rightarrow 0$). Since the Kramers-Kronig relations relate the real and imaginary parts of a response function, it is sufficient to compute either the real or the imaginary part of a given transverse component.
Hence, we choose to focus on
\begin{align}
	&\Re[\sigma_{xx}(\omega)] = \mathcal{C}_0 \, \sum_{\substack{n,\np, \\ s,\sp}} \, 
\int \, \frac{dk_z}{2\pi} \, 
 \frac{f_{n,s} - f_{\np,\sp}}{E_{n,s} - E_{\np,\sp}}  \, 
	\delta(\omega-E_{n,s}+E_{\np,\sp}) \, 
	\lrb 
	u_{n,s}^2 \, v_{\np,\sp}^2 \, \perm{\np}{J-1} \, \kr{n}{\np+1} +
	u_{\np,\sp}^2 \, v_{n,s}^2 \, \perm{\np}{J-1} \, \kr{n}{\np-1}
	\rrb \, ,
	\nn
&\Im[\sigma_{xy}(\omega)]= \mathcal{C}_0 \, \sum_{\substack{n,\np, \\ s,\sp}} \, 
	\int \, \frac{dk_z}{2\pi} \, 
 \frac{f_{n,s} - f_{\np,\sp}}{E_{n,s} - E_{\np,\sp}}  \, 
	\delta(\omega-E_{n,s}+E_{\np,\sp}) \, 
	\lrb 
	-u_{n,s}^2 \, v_{\np,\sp}^2 \, \perm{\np}{J-1} \, \kr{n}{\np+1} +
	u_{\np,\sp}^2 \, v_{n,s}^2 \, \perm{\np}{J-1} \, \kr{n}{\np-1}
	\rrb \, ,\nn
&  \mathcal{C}_0 = -(2^J \, e^2 \, J^2 \, \alpha_J^2) / (4\, l_B^{2J})\,,
\label{eq:mag_cond}
\end{align}
where the Dirac delta function implies energy conservation, and appears in the clean limit due to the identity
\begin{align}
	\lim\limits_{\epsilon \to 0} \, 
	\frac{1}{x+i \, \epsilon} = 
	\mc{P} \lrb\frac{1}{x} \rrb - i \, \pi \, \delta(x) \, .
\end{align}
Note that $\mc{P}$ stands for the principle value. 
We have provided a more simplified version of the expressions for $\Re[\sigma_{xx}]$ and $\Im[\sigma_{xy}]$ in Appendix~\ref{appx:simplify}.

\begin{figure}[] 
	\centering
	\includegraphics[width=\columnwidth]{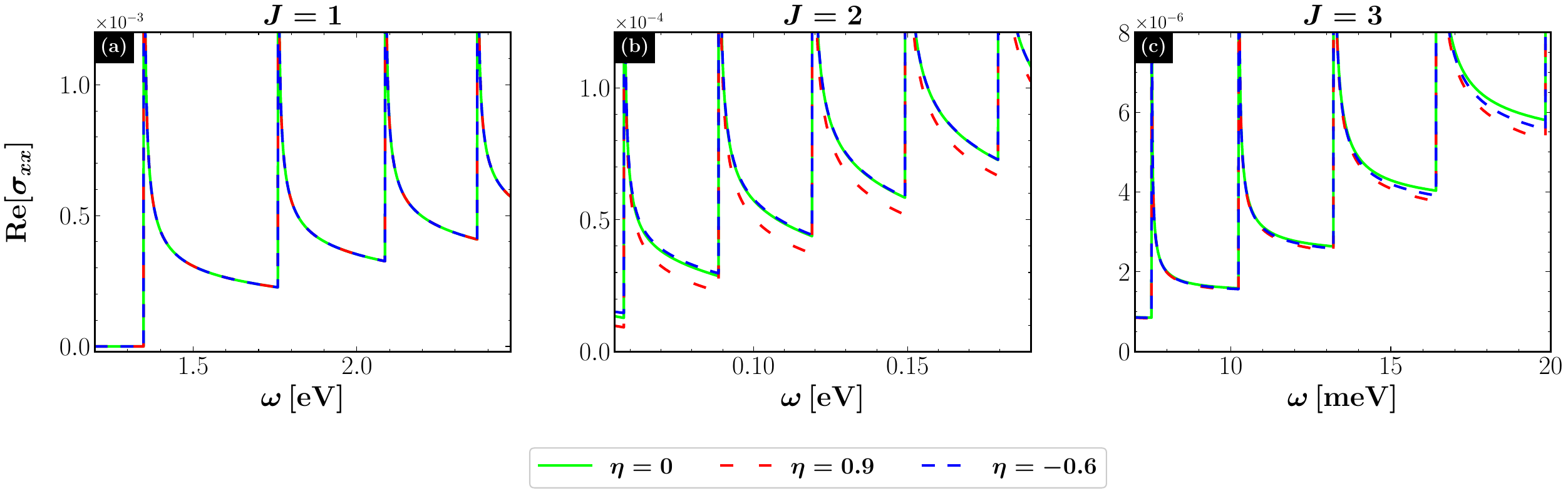}
\caption{The variation of the real part of $\sigma_{xx}$ [in units of $\ev^2$] as a function of the frequency $\omega$, for (a) WSM ($J=1$) with $\alpha_1 =0.5$, (b) double-WSM ($J=2$) with $\alpha_2 =1.2\times 10^{-1} \, \ev^{-1}$, and (c) triple-WSM ($J=3$) with $\alpha_3 =2.9\times 10^{-4} \, \ev^{-2}$, in the type-I phase without a cubic correction. We focus on the region $\omega > \lambda_J$ to show the response due to valence-to-conduction band transitions. The three curves in each subfigure correspond to the tilt parameter $\eta $ set to $0$ (green), $0.9$ (red), and $-0.6$ (blue), respectively. The remaining parameters are: $B=6.25 \, \ev^2$, $\beta=100 \, \ev^{-1}$, and $\mu=0 $.
\label{fig:cond_vc}}
\end{figure}

In the summations appearing in Eq.~\eqref{eq:mag_cond}, out of the four possible values of $(s,\sp)$, we note that $(\sp=-, \, s=+)$ refers to a transition from a valence band to a conduction band, while $(\sp= +, \, s= +)$ involves a transition between two conduction bands, via the absorption of a photon. Due to the energy-conservation from the Dirac delta function, a nonzero contribution is obtained only if 
\begin{align}
s\, \Delta_n (k_z) - \sp \Delta_{\np}(k_z) = \omega\, .
\label{eq:delta_cond1}
\end{align}
We multiply both sides of Eq.~\eqref{eq:delta_cond1} with $ \lsb s\, \Delta_n (k_z) + \sp \Delta_{\np}(k_z) \rsb$ to obtain
\begin{align}
\omega\left[	s\, \Delta_n (k_z) + \sp \Delta_{\np}(k_z) \right]
= 
\Delta_n^2 (k_z) - \Delta^2_{\np}(k_z)  
\Rightarrow
s\, \Delta_n (k_z) + \sp \Delta_{\np}(k_z) = 
	\frac{\lambda_J^2 \, \lrb\perm{n}{J} - \perm{n^{\prime}}{J} \rrb}{\omega} \, ,
\label{eq:delta_cond2}
\end{align}
using the value $\Delta_n(k_z) = \sqrt{g^2 (k_z) + \lambda_J^2 \, \perm{n}{J}}$ [cf. Eq.~\eqref{eqevs}].

Eqs.~\eqref{eq:delta_cond1} and~\eqref{eq:delta_cond2} lead to
\begin{align}
\label{eq:gkz}
g^2(k_z) = \lsb \frac{\lambda_J^2 \left (\perm{n}{J} - \perm{\np}{J} \right ) + \omega^2}
{2 \, \omega} \rsb^2 - \lambda_J^2 \, \perm{n}{J} \, .
\end{align}
The self-consistency of the equations puts a constraint on the upper bound of $n$, which
gives us the maximum value of the band-index of the Landau level allowed to participate in a transition process. This we define to be $\nmax$. Since $g(k_z)$ is a real function, its square should be non-negative. Now, this square is equal to the difference of two positive numbers, whose least value is zero.
Hence, we must have $g(k_z) \rightarrow 0$ in the limit $n \rightarrow \nmax$, leading to 
\begin{align}
	\lsb \frac{\lambda_J^2 
\left (\perm{n}{J} - \perm{\np}{J} \right) + \omega^2}
{2 \, \omega} \rsb^2  = \lambda_J^2 \, \perm{n}{J} \, ,
	\label{eq:nmax}
\end{align}
for $n =\nmax$. In generic cases, this has to be solved numerically in order to find the value of $\nmax$.

The constraint in Eq.~\eqref{eq:delta_cond1} also helps us estimate the location of peaks of the response function profile in the frequency space.
The constraints imposed by the Kronecker delta function in Eq.~\eqref{eq:m-matrix}, coupled with the condition in Eq.~\eqref{eq:nmax} [obtained from $g(k_z) \vert_{n= \nmax} \rightarrow 0$], tells us that
\begin{align}
\label{eqpeak}
	\omega \vert_{\rm{peak}} = 
	\lambda_J \, \lrb \sqrt{\perm{n+1}{J}} \pm \sqrt{\perm{n}{J}} \rrb \, .
\end{align}
While the plus sign [corresponding to ($\sp=-, \, s=+$)] indicates a transition from a valence band to a conduction band, the minus sign [corresponding to ($\sp=+, \, s=+$)] refers to a transition between two conduction bands. The expression suggests that $\lambda_J$ sets a natural energy scale for the peaks to appear 
\footnote{Of course the contributions from these transitions are modulated by the Fermi-Dirac distribution functions present in the integrals in Eq.~\eqref{eq:mag_cond}. In fact, these two factors compete with each other giving us the resulting magneto-optical conductivity.}.
Therefore, peaks due to valence-to-conduction band transitions will appear at $\omega >\lambda_J$. On the contrary, the peaks arising from the conduction-to-conduction band transitions appear at
\begin{align}
\label{eqpeak2}
\lim _{n \rightarrow \infty} \omega \vert_{\rm{peak}} = 
\begin{cases}
\lambda_1 \left( \sqrt{n+1} -\sqrt{n} \right) \rightarrow 0 &\text{ for } J=1 \\
\lambda_2 &\text{ for } J=2 \\
{3\, \lambda_3 \,\sqrt{n}} / {2}  &\text{ for } J=3 \\
	\end{cases} ,
\end{align}
in the large-$n$ limit (i.e., $n\gg 1$).
Based on these results, we will investigate relevant ranges of the spectrum to unveil the characteristics of various kinds of transitions, and we will also analyze how normal tilting or over-tilting affects this landscape.

\section{Numerical results}
\label{sec:result}

In this section, we showcase the numerical results for the magneto-optical conductivity, and discuss the effects of tilt. We would also like to point out that although the WSM case has already been studied in this context in Ref.~\cite{carbotte13,marcus-emil,mark_o,yu16}, we include it here for the sake of completeness and comparison.

\subsection{Type-I phase}
\label{sec:typeI}

Type-I phase (with or without a finite tilt) is characterized by point-like Fermi surfaces when the chemical potential cuts the band-crossing point. In our Hamiltonian, a type-I phase is realized for $|\eta|<1$. The dispersion for $\eta=0.7$ is shown in Fig.~\ref{fig:cond_cc}(a).

In order to understand the role of the tilt, we first note that the tilt parameter has no effect on the locations of the peaks, as shown in Eq.~\eqref{eqpeak}. Next, we need to check what happens to the response in the regions between consecutive peaks. For this, we need to focus on the difference of the Fermi-Dirac distribution factors [viz., $ \left( f_{n+1,s}-f_{n,s} \right)$] in Eq.~\eqref{eq:mag_cond}, as this would be the dominant factor to affect the behaviour due to tilt. This follows from the fact that the other terms, which involve $ \left \lbrace u_{n,s}, v_{n,s}, u_{n',s'}, v_{n',s'} \right \rbrace$, are independent of $\eta$, as these are functions of $\left \lbrace  E(k_z)-h(k_z) \right \rbrace $ [cf. Eq.~\eqref{eqestates}]. Additionally, the tilt terms cancel in factors such as $(E_{n,s}-E_{\np,\sp})$. Keeping the selection rules in mind, we find that
\begin{align}
& f_{n+1,s}-f_{n,\sp} 
= \frac{\mc{A} \, \lrb 
		e^{\beta \, \sp \Delta_n} - 
		e^{\beta s \,\Delta_{n+1}} \rrb}
	{1+ \mc{A} \, \lrb 
		e^{\beta \, \sp \Delta_n} + 
		e^{\beta s \,\Delta_{n+1}} \rrb + 
		\mc{A}^2 \, e^{\beta \, \lrb \sp \Delta_n + s \,\Delta_{n+1} \rrb} } \, ,\quad
 \mc{A} = e^{\beta \, h(k_z)}\,.		
	\label{eq:fermi_diff}
\end{align}
As shown in Fig.~\ref{fig:cond_cc}(a), the Landau levels are closest to the $\mu=0$ Fermi level around the $k_z=0$ region. This indicates that the transitions in the vicinity of $k_z=0$ will have the maximum impact on the conductivity. This allows us to reasonably assume $g(k_z) \approx 0$ for large contributions
\footnote{This follows from the fact that $g(k_z)$ is a cubic polynomial in $k_z$, with no $k_z$-independent term.
} --- \textit{a central result for simplifying the equations in the type-I phase}. We note that this condition also gives us the value of $\nmax$ [cf. Eq.~\eqref{eqpeak}].

\begin{figure}[] 
	\centering
	\includegraphics[width=\columnwidth]{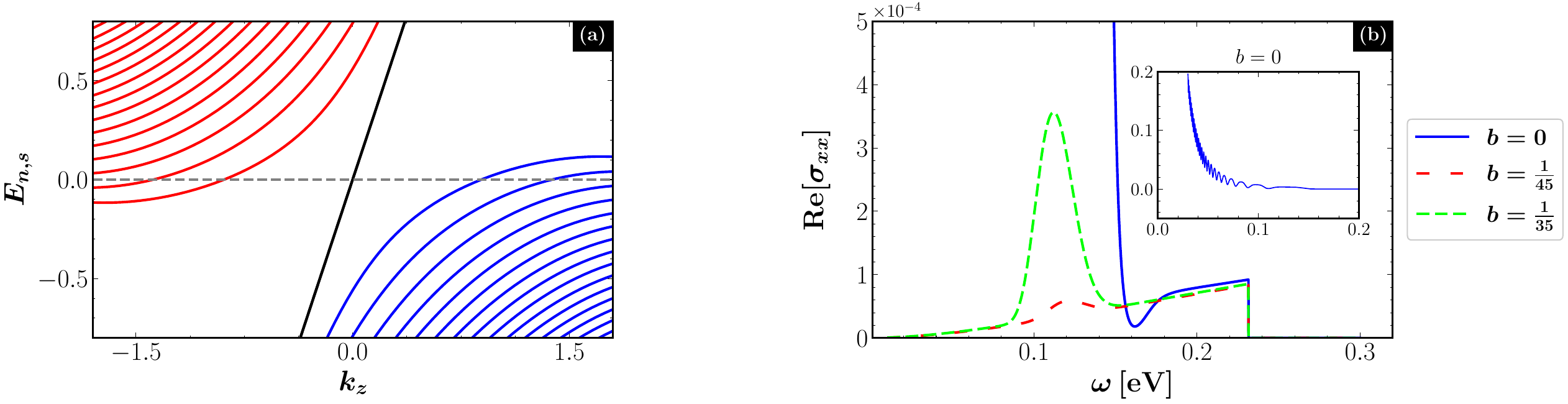}
	\caption{
Characteristics for a WSM ($J=1$) with $ \alpha_1 =0.5 $, in the type-II phase with $\eta=1.2$ [cf. Eq.~\eqref{eqevs}], when subjected to a magnetic field of magntitude $B=6.25 \, \ev^2$ at $\beta=100 \, \ev^{-1}$ (resulting in $\lambda_1 = 0.56\,\ev$).
Subfigure (a) shows the dispersion (in eV) of the first few Landau levels, as a function of the momentum $k_z$ (in eV), with the cubic correction coefficient $b=1/35$. The grey dashed line indicates the Fermi level, which is set to $\mu=0$. Subfigure (b) is a plot of the real part of $\sigma_{xx}$ [in units of $\ev^2$] as a function of frequency $\omega$, focussing on the region $ 0 <\omega < \lambda_1 $, in order to capture the contributions from conduction-to-conduction band transitions. Here, $\mu$ is set to zero. The three curves correspond to three values of $b$, as shown in the plotlegends. The inset shows a zoomed-out region for the $b=0$ case, highlighting the appearance of a series of peaks.}
	\label{fig:cond_cc_typeII}
\end{figure}

First, we focus on the conduction-to-conduction band transitions, for which $s=\sp= +$. Because both $\Delta_n$ and $\Delta_{n+1}$ are large for $n = \nmax$, the first two terms in the denominator of Eq.~\eqref{eq:fermi_diff} are much smaller than the third, and thus we can safely ignore them.
Hence, we can write Eq.~\eqref{eq:fermi_diff} in the large-$n$ limit as
\begin{align}
f_{n+1,+}-f_{n,+}  \approx e^{-\beta \left  (h+\Delta_{n+1} \right )} \, 
\lsb 1-e^{\beta \left (\Delta_{n+1}-\Delta_n \right )} \rsb  \, .
\end{align}
There is also the factor $E_{n+1, +} - E_{n,+} =\Delta_{n+1} -\Delta_n $ appearing in the denominator
of the integrands in Eq.~\eqref{eq:mag_cond}, which will affect the final results. Let us now analyze the scenarios on a case-by-case basis:
\begin{enumerate}

\item
WSM --- According to Eq.~\eqref{eqpeak}, the peaks exist near the zero-frequency region.
Ignoring $g(k_z)$, we find that the large-$n$ limit yields $(\Delta_{n+1}-\Delta_n) \approx {\lambda_1} / {\sqrt{n}}$ for $J=1$. Therefore, $\left( f_{n+1,+}-f_{n,+}  \right)
/ \left (\Delta_{n+1}-\Delta_n \right )
\approx e^{-\beta \left (h+\Delta_{n+1} \right )} 
\,  \beta  $, which leads to an exponential suppression of the peaks in the small frequency regimes (not shown in the plots).

\item
double-WSM --- According to Eq.~\eqref{eqpeak}, the peaks appear in the $\omega \sim \lambda_2 $ region.
Setting $J=2$ yields $(\Delta_{n+1}-\Delta_n) \approx \lambda_2 $ for large $n$. 
This leads to well-defined peaks in the $\omega \sim \lambda_2 $ region (with no exponential suppression).
Since $\Delta_{n+1} \gg h(k_z)$, we find that the dependence on tilt remains negligible. Since the gap between consecutive Landau levels near the $k_z = 0$ tends to $\lambda_2$ [as seen in Fig.~\ref{fig:cond_cc}(a)], there are increasingly more and more transitions possible as we move closer and closer to $\lambda_2$. In fact, the conductivity is unbounded (i.e., goes to infinity), as is clearly seen in Fig.~\ref{fig:cond_cc}(b), when we move to the $\omega =\lambda_2$ point from higher values of $\omega$.
These peaks have not been reported in the earlier studies.

\item
triple-WSM --- 
The peaks appear far away from $\omega \sim \lambda_3 $, as explained in the discussions following Eq.~\eqref{eqpeak}. Hence, we do not show the results for $J=3$ here.

\end{enumerate}

The second case involves the valence-to-conduction band transitions, for which we have $\sp=-$ and $s=+$.
Let us again analyze the scenarios on a case-by-case basis: 
\begin{enumerate}

\item
WSM --- 
As both $\Delta_{n+1}$ and $\Delta_n$ become large at $ n = \nmax$, terms like $e^{\beta \, \sp \Delta_n}$ vanish in Eq.~\eqref{eq:fermi_diff}, which allows us to approximate
\begin{align}
f_{n+1,1}-f_{n,-1}  \approx 
	-\frac{\mc{A} \, e^{\beta \, \Delta_{n+1}}}
	{1+ \mc{A} \, e^{\beta \, \Delta_{n+1}}} =-1 \,.
\end{align}
Since the above expression is independent of the tilt factor, the $J=1$ conductivity should be independent of $\eta$. This is in agreement with the results shown in Fig.~\ref{fig:cond_vc}(a), which also agrees with the results in Ref.~\cite{mark_o}. 

\item
double-WSM --- For $J=2$, we get
\begin{align}
f_{n+1,1}-f_{n,-1}  \approx - \frac{e^{\beta \, \lambda_2 \, n} \, (e^{\lambda_2} -1 )}
{1- \mc{A} \, (e^{\lambda_2} -1 )} \, ,
\end{align}
where we have again used the result $(\Delta_{n+1}-\Delta_n) \approx \lambda_2 $ for the large-$n$ limit.
Clearly, $\mc{A}$ is negligible for a negative tilt ($-1<\eta<0$). A positive tilt ($0<\eta <1 $) makes $\mc{A}$ large, which in turn reduces the value of $\left( f_{n+1,1}-f_{n,-1} \right)$, leading to a suppression of the conductivity. As shown in Fig.~\ref{fig:cond_vc}(b), such suppression takes place in the ``tail'' region (i.e., the region between consecutive peaks) of the conductivity for a positive tilt, and disappears for either zero or negative tilt. 

\item
triple-WSM --- Since the complexity of the equations increases many-fold with increasing $J$, a closed-form analytical treatment is not possible for the $J=3$ case. Fig.~\ref{fig:cond_vc}(c) shows the dependence on tilt for a triple-WSM. The behaviour of the conductivity curves shows that the suppression of the optical conductivity is prominent only at higher frequencies.

\end{enumerate}

\begin{figure}[] 
	\centering
	\includegraphics[width=\columnwidth]{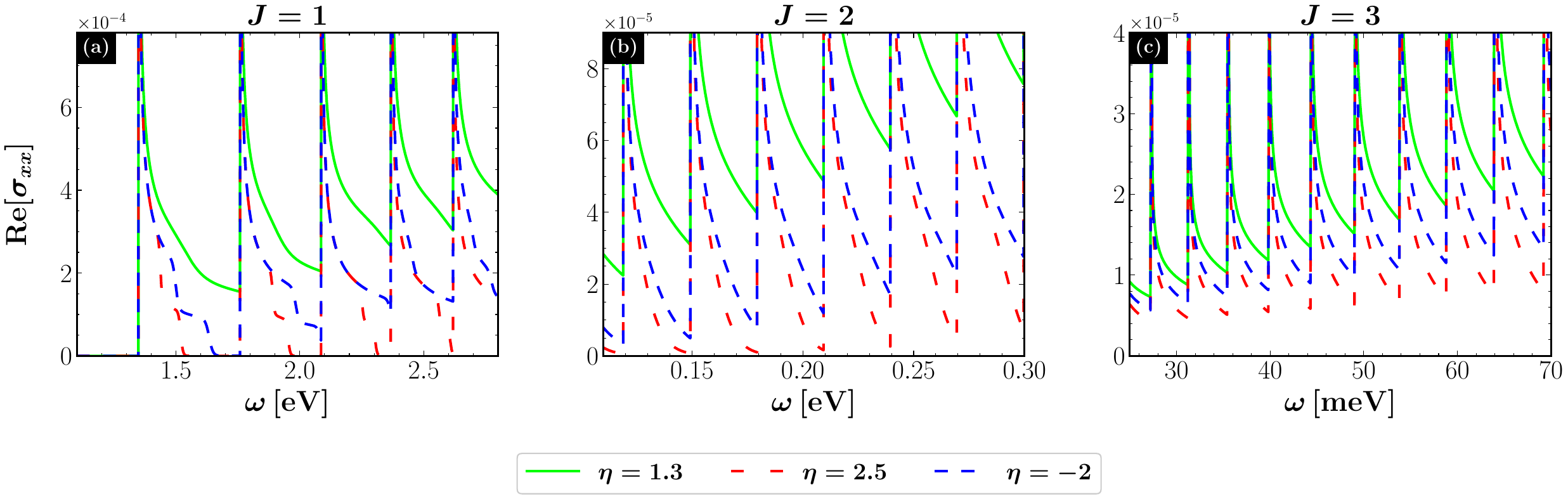}
\caption{
The variation of the real part of $\sigma_{xx}$ [in units of $\ev^2$] as a function of the frequency $\omega$, for (a) WSM ($J=1$) with $\alpha_1 =0.5$, (b) double-WSM ($J=2$) with $\alpha_2 =1.2\times 10^{-1} \, \ev^{-1}$, and (c) triple-WSM ($J=3$) with $\alpha_3 =2.9\times 10^{-4} \, \ev^{-2}$, in the type-II phase. We focus on the region $\omega > \lambda_J$ to show the response due to valence-to-conduction band transitions. The three curves in each subfigure correspond to the tilt parameter $\eta $ set to $1.3$ (green), $2.5$ (red), and $-2$ (blue), respectively. The remaining parameters are: $b=0$, $B=6.25 \, \ev^2$, $\beta=100 \, \ev^{-1}$, and $\mu=0 $.
\label{fig:cond_vc_typeII1}}
\end{figure}

\subsection{Type-II phase}
\label{sec:typeII}

The type-II phase is realized by restricting to $|\eta|>1$, which makes the cone over-tilted, leading to open Fermi surfaces. To address the issue of the unphysical case of an open Fermi surface, one may need to use either a hard cutoff, or higher-order curvature terms in the dispersion. We take the second route by using a nonzero $b$.

First, we determine the nature of the conduction-to-conduction band transitions captured by $s=\sp=+$.
Over-tilting allows new transitions which were forbidden for the type-I case, and the restriction of the nonzero contributions to the $k_z \simeq 0$ region is no longer applicable. For a given Landau level $n$ and a frequency $\omega$, we can evaluate the relevant $k_z$ using Eq.~\eqref{eq:delta_cond1}. This is obtained by solving
\begin{align}
g^2(k_z) &= \left[ \frac{\lambda_J}{ 2 \, \omega } (\Delta_{n+1} 
- \Delta_{n}) + \frac{\omega}{ 2 \, \lambda_J} \right]^2 - \Delta_{n+1} \,.
\label{eq:overtilt_cond}
\end{align}
Let us now analyze the scenarios on a case-by-case basis:

\begin{enumerate}

\item
WSM --- Because of over-tilting, the Fermi level at $\mu =0$ can now cut Landau levels away from the $k_z=0$ region (in addition to the vicinity around $k_z = 0$), as seen in Fig.~\ref{fig:cond_cc_typeII}(a). This unlocks new transitions that are absent in the type-I phase, thus significantly affecting the transport properties. For a single value of $\omega$, Eq.~\eqref{eq:overtilt_cond} is satisfied for multiple values of $n$ and $k_z$, giving rise to multiple peaks at one single frequency. Such contributions keep on increasing as $\omega$ decreases, overpowering the suppressing effect of the Fermi-Dirac distribution factors shown in Eq.~\eqref{eq:fermi_diff}. This leads to an unbounded growth of conductivity for $b=0$ (open Fermi pockets), as shown in Fig.~\ref{fig:cond_cc_typeII}(b). To elaborate further, $\Re[\sigma_{xx}(\omega)]$ is found to be nonzero [cf. Fig.~\ref{fig:cond_cc_typeII}(b)] even at a frequency close to $\lambda_1$, in contrast with the type-I case. The inset of Fig.~\ref{fig:cond_cc_typeII}(b) magnifies the behaviour when cubic correction is zero, where we find that the $\Re[\sigma_{xx}(\omega)]$ profile supports numerous small peaks arising from the conduction-to-conduction band transitions, superimposed on a sharply decaying (with increasing $\omega$) background. The density of these tiny peaks increases rapidly as one approaches zero frequency (accompanied by a reduction in peak height with respect to the background). Their origin is attributed to an open Fermi surface, because for a realistic scenario with a nonzero cubic correction $b=1/45$, these features get replaced by a single well-pronounced peak. As $b$ is slightly increased to the value $1/35$, the height f this single peak increases many-fold. 

\item
double-WSM --- For $J=2$, we have already established the existence of well-pronounced peaks (due to conduction-to-conduction band transitions) in the neighbourhood of $\omega \sim \lambda_2$ in the type-I phase [cf. Fig.~\ref{fig:cond_cc}(b)]. Over-tilting the cone continues to demonstrate the same features. Extra transitions unlocked due to over-tilting can only give rise to new peaks for small values of $\omega$, because the Landau levels get denser and denser as we move away from $k_z = 0$. As a result, they are cloaked by the peaks shown in Fig.~\ref{fig:cond_cc}(b), which are not the artifacts of open Fermi pockets. Hence, the cubic correction is incapable of curing the divergent conductivity near $\omega = \lambda_2$ (discussed in Sec.~\ref{sec:typeI}).

\item
triple-WSM --- For $J=3 $, Eq.~\eqref{eq:overtilt_cond} can be simplified using the fact that $\omega$ is small for new transitions resulting from over-tilting. Here, we find that there is only one value of $\lbrace k_z,n \rbrace $ for a given frequency. The Fermi-Dirac distribution factors shown in Eq.~\eqref{eq:fermi_diff} suppress this contribution, leading to negligible conductivity, because there is not enough number of transitions to overcome this suppression. This implies that there is no need for a hard momentum cutoff or a cubic correction. There are still transitions near $k_z = 0$, carrying forward from the type-I scenario, but the corresponding peaks are located far away from $\lambda_3$. Hence, we have not shown the corresponding results.

\end{enumerate}

For the valence-to-conduction band transitions, featured in the $\omega $-regions away from $\lambda_J$, we find that over-tilting of the cone strongly suppresses the tails in the $\Re[\sigma_{xx}(\omega)]$-profile, compared to the type-I case, irrespective of the value of $J$. This is shown in Fig.~\ref{fig:cond_vc_typeII1}. Interestingly, $\Re[\sigma_{xx}(\omega)]$ for $J=1$ in the type-II phase (for both positive and negative tilt) hosts prominent ``plateau-like'' features, whose width increases at lower frequencies. These features are absent in the multi-WSMs. 

\begin{figure}[] 
\centering
\includegraphics[width=\columnwidth]{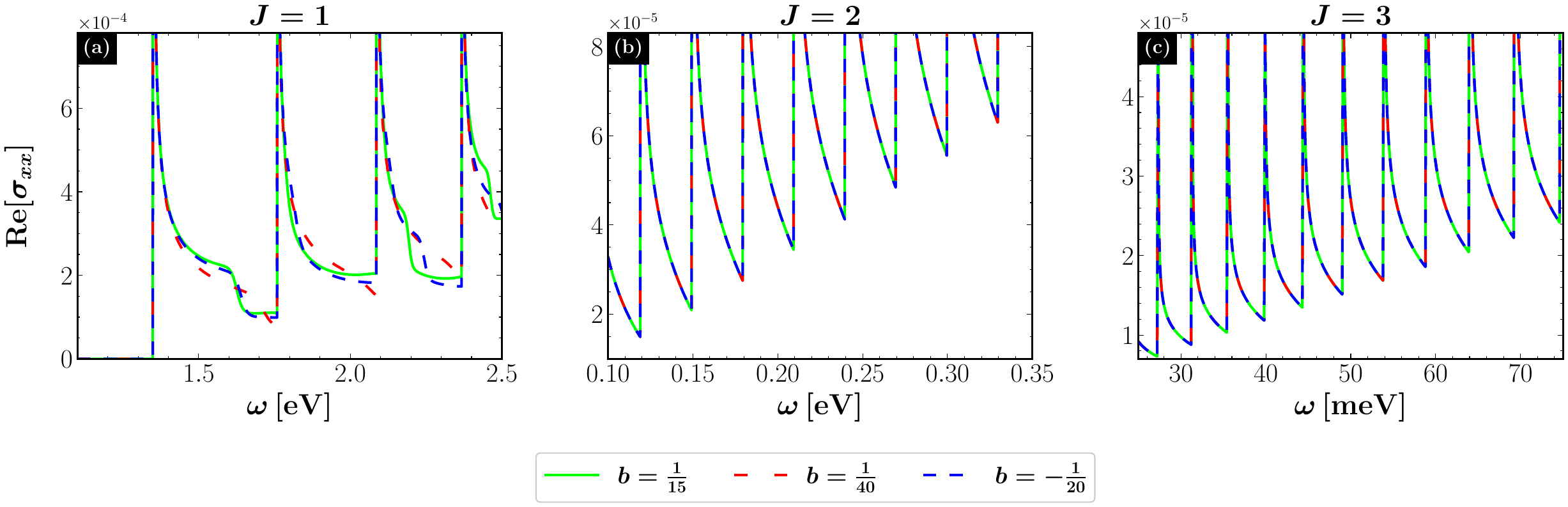}
\caption{The variation of the real part of $\sigma_{xx}$ [in units of $\ev^2$] as a function of the frequency $\omega$, for (a) WSM ($J=1$) with $\alpha_1 =0.5$, (b) double-WSM ($J=2$) with $\alpha_2 =1.2\times 10^{-1} \, \ev^{-1}$, and (c) triple-WSM ($J=3$) with $\alpha_3 =2.9\times 10^{-4} \, \ev^{-2}$, in the type-II phase. We focus on the region $\omega \gg \lambda_J$ to show the response due to valence-to-conduction band transitions. The three curves in each subfigure correspond to the cubic correction coefficient $ b $ set to $1/15$ (green), $1/40 $ (red), and $-1/20$ (blue), respectively.
The remaining parameters are: $\eta=1.5$, $B=6.25 \, \ev^2$, $\beta=100 \, \ev^{-1}$, and $\mu=0 $.
\label{fig:cond_vc_typeII2}}
\end{figure}

In Fig.~\ref{fig:cond_vc_typeII2}, we show the effects of nonzero cubic corrections on the peaks originating from the valence-to-conduction band channels. The curves include the case with a negative $b$, which corresponds to the scenario that open Fermi pockets exist in this case, as the sign of the curvature is such that it does not close the Fermi pockets. However, it has negligible effects on the conductivity curves, plotted using the data from our numerical simulations. From the initial expressions of the conductivity components [cf. Eq.~\eqref{eq:mag_cond}], we can infer that this insensitivity is due to the higher-momentum regions being prevented from causing divergences in the integrals (by the Dirac delta functions), within the frequency ranges under consideration here (which involve $\omega >\lambda_J$). This also justifies why we have used the value $b=0$ in many of our plots in the type-II phase (e.g., in Fig.~\ref{fig:cond_vc_typeII1}).

In addition to the linear tilt, one can also consider a quadratic tilt~\cite{marcus-emil}, for example, parametrized by $ \eta' \,  k_z^2$. Due to the higher power of momentum appearing in the expression, the energy gaps (due to the $ \eta' \,  k_z^2$ term) away from $k_z=0$ become smaller compared to those obtained for a purely linear tilt. Thus, when we consider the transitions in the presence of a Fermi pocket, taking place away from the $k_z=0$ region, the peak frequencies corresponding to these tiny energy gaps are also extremely small. It is possible to study these new peaks for WSMs, but for $J=2$, the peaks get overshadowed by the series of large peaks, that are already present in the conductivity spectrum. Finally, for the case of $J=3$, the equations are too complicated for an analytical treatment, and from our numerical simulations (not shown here), we do not see any significant effects for either $\omega < \lambda_3$ or $\omega > \lambda_3$.

\subsection{Circularly polarized magneto-optical conductivity}
\label{sec:pol}

\begin{figure}[] 
	\centering
	\includegraphics[width=\columnwidth]{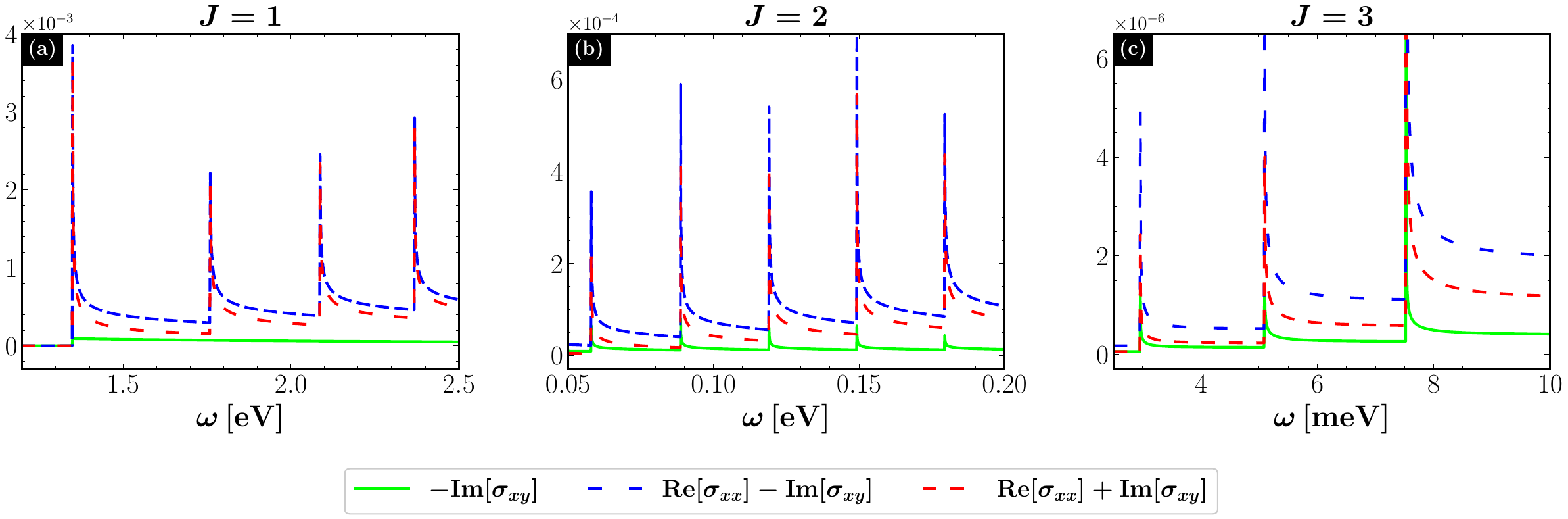}
	\caption{
The variation of various combinations of the conductivity tensor components [in units of $\ev^2$] as functions of the frequency $\omega$, for (a) WSM ($J=1$) with $\alpha_1 =0.5$, (b) double-WSM ($J=2$) with $\alpha_2 =1.2\times 10^{-1} \, \ev^{-1}$, and (c) triple-WSM ($J=3$) with $\alpha_3 =2.9\times 10^{-4} \, \ev^{-2}$, in the absence of any tilt or cubic correction. The three curves in each subfigure correspond to $ \Re[\sigma_{+}]$ (blue), $ \Re[\sigma_{-}]$ (red), and $ -\Im [\sigma_{xy}]$ (green), respectively.
The remaining parameters are: $B=6.25 \, \ev^2$, $\beta=100 \, \ev^{-1}$, and $\mu=0 $.
\label{fig:cond_pol}}
\end{figure}

Till now, we have only demonstrated the behaviour of the diagonal part $ \sigma_{xx} $ ($=  \sigma_{yy} $) of the transverse components of $\sigma_{\mu \nu}$, which gives us insights into the bandstructure and the Landau levels. On the other hand, the off-diagonal parts, $\sigma_{xy} =\sigma_{yx}$, in conjunction with $\sigma_{xx}$, determine the behaviour for circularly polarized light~\cite{bordacs11,carbotte13}. 
More specifically, in experiments that probe the circular polarization of resonant light, as in the case of the
Kerr and Faraday effects, the absorptive parts of the conductivity are given by the combinations $\sigma_{\pm} = \sigma_{xx} \pm i \,  \sigma_{xy} $, where the ``$+$''(``$-$'') sign corresponds to the right(left)-handed polarization. In Fig.~\ref{fig:cond_pol}, we have plotted $\Re[\sigma_{\pm}] = \Re[\sigma_{xx}]\mp \Im[\sigma_{xy}]$ against frequency for all the three $J$-values. In all cases, we find that the spectrum consists of multiple peaks, with the magnitudes of the peaks for $\Re[\sigma_{+}]$ being larger than those for $ \Re[\sigma_{-}]$. The value of $J$ affects the location and the density of the peaks, as well as the overall magnitude of the conductivity. 

While $\Re[\sigma_{\pm}]$ provides information about a specific polarization of light, it is easy to check that the absorptive part of the off-diagonal conductivity $ \sigma_{xy} =  \sigma_{yx} $, given by $\Im[\sigma_{xy}] = 
\lrb \Re[\sigma_{-}]-\Re[\sigma_{+}] \rrb/2$, captures the difference in the conductivity response for the two opposite polarizations, and is proportional to the power absorption spectrum. Hence, in Fig.~\ref{fig:cond_pol}, we have also plotted $ -\Im[\sigma_{xy}]$ as a function of frequency, which again shows a strong dependence on the specific $J$-value.

\section{Summary and outlook}
\label{sec:conclusion}

In this paper, we have investigated the magneto-optical conductivity of Weyl and multi-Weyl semimetals. An external magnetic field gives rise to Landau levels, which in turn lead to a series of asymmetric peaks in the real part of the component $\sigma_{xx}$. 
We have shown that a natural energy scale $\lambda_J \propto B^{J/2}$ emerges for the existence of these peaks, implying that the WSMs and multi-WSMs with unequal topological charges host the peaks at different frequencies. This is a more generic result as it explains the previously-observed $\sqrt{B}$-scaling in graphene and WSMs~\cite{sharapov-graphene,carbotte13,magneto-double-weyl,mark_o,marcus-emil}.
We also note that a two-dimensional slice of a multi-WSM at $k_z=0$ has a dispersion identical to certain stacked graphene multilayers with interlayer tunneling. This is because, such a system has a dispersion proportional to $|\mathbf k|^{J/2}$ --- in the presence of a perpendicular magnetic field of magnitude $B$, the Landau levels, which have nonzero energy, show a dependence proportional to $B^{J/2}$~\cite{min_allan, min_allan_2}. Hence, the peak positions in the magneto-optical conductivity of the multi-WSMs show similarity to these multilayered graphene systems.

Our calculations reveal that the contribution to a peak comes from either valence-to-conduction band transitions, or conduction-to-conduction transitions. The peaks due to the former processes mostly appear far away from $\lambda_J$, while the peaks for the latter case emerge at diverse frequency ranges, depending on $J$. 
As a crucial part of our investigations, we have demonstrated how tilting the cone non-trivially affects the conductivity peaks. In particular, for the type-I phases (with tilt parameter $ |\eta| <1$), the transitions with leading contributions occur close to $k_z=0$. This feature has allowed us to make a reasonable approximation, which elucidates the connection between tilt and conductivity. Our results show that the type-I conductivity of multi-WSMs (for valence-to-conduction band channels) is suppressed for frequencies away from $\lambda_J$, a phenomenon non-existent in WSMs. We have also noticed that such a suppression appears to be frequency-dependent for triple-WSMs, because the suppression is prominent only at higher frequencies. The conductivity due to the conduction-to-conduction band channels is finite only for multi-WSMs (i.e., for $J>1$). 

In the type-II phase, the over-tilting of the cone produces Fermi pockets in the dispersion. Since unbounded Fermi pockets may necessitate the introduction of cutoffs in the momentum integrals, we have supplemented the linear-dispersion in the $k_z$-direction with a cubic correction. The possible divergences are likely to appear only around the regime $\omega\sim \lambda_J$ [as explained in Eq.~\eqref{eqpeak}], where conduction-to-conduction band transitions are activated. But ultimately, the cubic correction coefficient $b$ does not seem to play any significant role, except in the type-II phase of the WSMs, as described below.

Unlike the type-I phase, the WSM conductivity close to $\omega= \lambda_1$ is not zero, and in fact, it exhibits a series of dense peaks when the cubic correction is set to zero. This can be attributed to the newly allowed transitions, which are forbidden in the absence of over-tilting. For a finite cubic correction, the series of peaks in the $J=1$ spectrum disappears, and gets replaced by a single peak. For $J=2$, the conductivity diverges as $\omega \rightarrow \lambda_2$, both in the type-I and type-II phases, which veils the small new peaks due to over-tilting --- this tells us that a hard momentum cut-off or a cubic correction cannot cure this divergence, which is an inherent feature of the double-WSM case irrespective of the tilt. New peaks due to over-tilting do not appear for $J=3$, because the Fermi-Dirac distribution factors in the integrand kill the terms arising from the extra transition-terms. This makes $b$ redundant also in the $J=3$ case. Away from $\lambda_J$, we have found that the conductivity is suppressed if the tilt is switched from type-I to type-II phase, irrespective of the value of $J$. 

Finally, we have demonstrated the characteristics of the circularly polarized conductivity, which comes in two different linear combinations, viz., $\Re[\sigma_{\pm }] = \Re[\sigma_{xx}] \mp \Im[\sigma_{xy}]$. Interestingly, the absorptive part of the conductivity for a left-handed polarized light (i.e., $\Re[\sigma_+]$) is larger in magnitude than that for the right-handed counterpart (i.e., $\Re[\sigma_-]$). As a result, the power absorption spectrum, which is proportional to the difference $2 \Im[\sigma_{xy}]=  \Re[\sigma_{-}]-\Re[\sigma_{+}] $, is nonzero. This could potentially be used as a signature for distinguishing semimetals with different topological charges, similar to the response related to circular dichroism~\cite{goldman_dir,sajid_cd}. For all these physical observables, the exact features are dependent on $J$.

In summary, our magneto-optical conductivity results, being accessible by current experimental techniques, serve to provide transport characteristics for WSMs and multi-WSMs, complementing other possible signatures~\cite{dantas,ipsita-kush,Deng2020,ipsita-aritra,nag2020thermoelectric,sajid_cd,ipsita-sajid,ipsita-serena,ips-sandip-sajid}. Our findings supplement earlier studies of magneto-optical conductivity for various semimetals (which include WSMs and nodal-line semimetals~\cite{carbotte13,magneto-double-weyl,mark_o,marcus-emil,nodal-line, nodal-line-type2}), especially by including the effects of higher powers of momentum in the dispersions, included as corrections
to the leading-order low-energy effective models. In future, it will be worthwhile to consider such higher-order curvature terms in the context of computing the magneto-optical conductivity tensor for the type-II phases of nodal-line semimetals. Another interesting direction will be to study the magneto-optical conductivity in the presence of interactions and/or impurities~\cite{rahul-sid,ipsita-rahul,ips-qbt-sc,ips-biref,ips-klaus}.
This will make the scattering rate $\epsilon$ [cf. Eq.~\eqref{eqkubo}] highly nontrivial, and it has to be computed via fermion self-energy. Furthermore, very strong interactions may induce many-body effects like (i) destroying the quantization of various physical quantities, that are topologically protected in the topological phases of the single-particle Hamiltonians \cite{kozii,ips-photocurrent}; (ii) emergence of strongly correlated phases~\cite{ips-seb,MoonXuKimBalents,rahul-sid,ipsita-rahul,ips-qbt-sc,ips-biref}, where the quasiparticle picture for deriving semiclassical transport equations breaks down \cite{ips-subir,ipsc2,ips-mem-mat,ips-freire2,ips-freire3,zero-sound}. Addressing such scenarios will necessitate the use of many-body techniques like Green's function method and Keldysh formalism \cite{zero-sound}.

\section*{Acknowledgments}
We are grateful to Marcus St\aa{}lhammar for insightful discussions. We also thank Ahmed Bouhlal and Ahmed Jellal for participating in the rudimentary stages of this project. S.S. is funded by the National Science Centre (Narodowe Centrum Nauki), Poland, under the scheme Preludium Bis-2
(Grant Number 2020/39/O/ST3/00973).

\appendix

\section{Landau levels for Weyl and multi-Weyl semimetals}
\label{appx:eigensystem}

In this appendix, we outline the derivation of the eigensystem for the Hamiltonian in Eq.~\eqref{eq:ham2}.
The corresponding eigenvalue equation can be written as
\begin{align}
\pmtx{h+g && \lambda_J \, a^J 
		\\ \lambda_J \, (a^{\dagger})^J &&  h-g} \,
	\ket{\Psi_{n,s}} =
	E_{n,s} \, \ket{\Psi_{n,s}} \, ,
\end{align} 
where we have suppressed the $k_z$-dependence. Eq.~\eqref{eqestates} tells us that the eigenstates for the non-degenerate ($n\ge J$) bands are given by $\ket{\Psi_{n,s}} = \pmtx{s\, \mc{U}_{n,s} && \mc{V}_{n,s}}^T$. Hence, the eigenvalue equation gives us
\begin{align}
	\lrb h + g \rrb s\, \mc{U}_{n,s} + 
	\lambda_J \, a^J \mc{V}_{n,s} = s\, E_{n,s}\, \mc{U}_{n,s}\, , \quad
	\lrb h - g \rrb \mc{V}_{n,s} + 
	s\, \lambda_J \, (\adag)^J \, \mc{U}_{n,s} = E_{n,s}\, \mc{V}_{n,s}\, .
\end{align}
The above equations can be decoupled by using the familiar trick of
multiplying each one by appropriate factors of $a$ and $a^\dagger$. This leads to
\begin{align}
	a^J \, (\adag)^J \, \mc{U}_{n,s}= \frac{(E_{n,s}-h)^2-g^2}{\lambda_J^2} \, \mc{U}_{n,s}\, , \quad
	(\adag)^J \, a^J\, \mc{V}_{n,s}= \frac{(E_{n,s}-h)^2-g^2}{\lambda_J^2} \, \mc{V}_{n,s}\, .
	\label{eq:a3}
\end{align}

From the structure of the equations, we make the ansatz $\mc{U}_{n,s} = u_{n,s} \, \Phi_{n-J}$ and $\mc{V}_{n,s} = v_{n,s} \, \Phi_{n}$, where $\Phi_n$ is the $n^\t{th}$ eigenstate of a quantum harmonic oscillator, satisfying the relations $a\, \Phi_n = \sqrt{n}\, \Phi_{n-1}$ and $\adag\, \Phi_n = \sqrt{n+1}\, \Phi_{n+1}$.
Plugging this in Eq.~\eqref{eq:a3}, we find that
\begin{align}
	E_{n,s} = \begin{cases}
h+g  & \text{ for } n = 0, 1, \cdots , J-1 \\
h + s\, \sqrt{g^2+ \lambda_J^2 \, \perm{n}{J-1}} & \text{ for } n \geq J 
\end{cases}	\, ,
\label{eq:a4}
\end{align}
where $\perm{n}{J}=n!/(n-J)!$. Making use of Eq.~\eqref{eq:a3} once more, we
obtain
\begin{align}
& \lambda_J \, \sqrt{\perm{n}{J}} \, u_{n,s} = 
s\, (E_{n,s}-h+g) \, v_{n,s} \,,\quad
\lambda_J \, \sqrt{\perm{n}{J}} \, v_{n,s} 
= s\, (E_{n,s}-h-g) \, u_{n,s} \, ,\nn
& \Rightarrow	\lrb 
\frac{v_{n,s}}{u_{n,s}} \rrb^2 = \frac{E_{n,s}-h+g}
{ E_{n,s}-h-g } \, .
\end{align}
Finally, imposing the orthonormality condition, the closed-form analytical expressions turn out to be
\begin{align}
\label{equv}
u_{n,s}= \begin{cases}
0 & \text{ for } n=0,1, \cdots, (J-1) \\
\sqrt{\frac{1}{2} \lrb 1+ \frac{g}{E_{n,s}-h} \rrb} & \text{ for } n\geq J
\end{cases}\,,
\quad
v_{n,s}= \begin{cases}
0 & \text{ for } n=0,1,\cdots, (J-1) \\
\sqrt{\frac{1}{2} \lrb 1- \frac{g}{E_{n,s}-h} \rrb} & \text{ for } n\geq J  
\end{cases} \,.
\end{align}

\section{Simplified expression for $\Re[\sigma_{xx}]$}
\label{appx:simplify}

In this appendix, we provide simplified expressions for $\Re[\sigma_{xx}]$ and $ \Im[\sigma_{xy}]$ derived in Sec.~\ref{sec:cond}.
From the expressions in Eq.~\eqref{eq:mag_cond}, we can easily identify that there are two independent terms whose linear combinations produce both $\Re[\sigma_{xx}]$ and $ \Im[\sigma_{xy}]$. Let us call these ${\mathcal S}_1$ and ${\mathcal S}_2$, such that $\Re[\sigma_{xx}] = {\mathcal S}_1 + {\mathcal S}_2$ and $\Im[\sigma_{xy}] = -{\mathcal S}_1 + {\mathcal S}_2$. The explicit expressions for these functions are then given by
\begin{align}
& {\mathcal S}_1 (\omega)
= {\mathcal C}_0 \, \sum_{\substack{n,\np, \\ s,\sp}} 
	\, \int \, \frac{dk_z}{2\pi} \,
\frac{f_{n,s} - f_{\np,\sp}}{E_{n,s} - E_{\np,\sp}}  \,
	\delta(\omega-E_{n,s}+E_{\np,\sp}) \, 
u_{n,s}^2 \, v_{\np,\sp}^2 \, \perm{\np}{J-1} \, \kr{n}{\np+1}  \, ,\nn
&{\mathcal S}_2  (\omega)
= {\mathcal C}_0 \, \sum_{\substack{n,\np, \\ s,\sp}}
	\, \int \, \frac{dk_z}{2\pi} \,
\frac{f_{n,s} - f_{\np,\sp}}{E_{n,s} - E_{\np,\sp}}  \,
	\delta(\omega-E_{n,s}+E_{\np,\sp}) \, u_{\np,\sp}^2 \, 
v_{n,s}^2 \, \perm{\np}{J-1} \, \kr{n}{\np-1} \, .
\label{eq:sigma12}
\end{align}

In accordance with the selection rule $\np=n\pm 1$ implemented by the Kronecker delta functions, the summation over $s$ and $\sp$ leads to the following possibilities (arising from the Dirac delta function constraints):
\begin{align}
	& \omega - E_{n,+} + E_{n+1,+}  = \omega - E_{n+1,-} + E_{n,-1}  = 0\,,\quad
\omega - E_{n,-} + E_{n+1,-}  = \omega - E_{n+1,+} + E_{n,+}  = 0\,, \nn
& \omega - E_{n,+} + E_{n+1,-}  = \omega - E_{n+1,-} + E_{n,-}  = 0\,, \quad
\omega - E_{n,-} + E_{n+1,+} = \omega - E_{n+1,+} + E_{n,+}  = 0\, ,
\end{align}
at least one of which must be satisfied to obtain a nonzero contribution to the integrals above.
Since we are only interested in the absorption spectra, we can drop the equations which are only satisfied for $\omega < 0$. This reduces the number of possibilities to two, which are given by
\begin{align}
\omega - E_{n,-} + E_{n+1,-} = \omega - E_{n+1,+} + E_{n,+} = 0\,, \quad
\omega - E_{n,+} + E_{n+1,-} = \omega - E_{n+1,-} + E_{n,-} = 0\,,
	\label{eq:reqonE}
\end{align}
provided we restrict ourselves to the $\omega \geq 0$ regions.
Since $E_{n,s} = h + s\, \Delta_n$, the first equation yields
\begin{align}
	\omega = \Delta_{n+1}-\Delta_n = 
	\sqrt{g^2 + \lambda_J^2 \, \perm{n+1}{J}} -
	\sqrt{g^2 + \lambda_J^2 \, \perm{n}{J}}\,.
\end{align}
Substituting the expression of $g$ from Eq.~\eqref{eq:gkz}, we obtain
\begin{align}
	\omega = 
	\frac{\lambda_J^2 \, \dperm{n}{J} + \omega^2}{2 \, |\omega|} -
	\frac{\big|\lambda_J^2 \, \dperm{n}{J} - \omega^2 \big|}{2 \, |\omega|} \, ,
\quad \dperm{n}{J} = \perm{n+1}{J} - \perm{n}{J} \,.	
\end{align}
Since we have already assumed that $\omega \geq 0$, the above equation is satisfied only if $\lambda_J^2 \, \dperm{n}{J} > \omega^2$, and thus can be replaced by the factor
$\Theta(\lambda_J^2 \, \dperm{n}{J} - \omega^2)$, with $\Theta$ being the Heaviside theta function. Following a similar argument, the second relation in Eq.~\eqref{eq:reqonE} is equivalent to $\Theta(\omega^2 - \lambda_J^2 \, \dperm{n}{J})$. Finally, the integration over $k_z$ can be simplified using the delta function property
\begin{align}
	\int \, f(x) \, \delta \big(\gamma (x)\big ) \, dx 
= \sum_{x_0} \, \frac{f(x_0)} { |\gamma^{\prime}(x_0)| }\, ,
\end{align}
where $\lbrace x_0 \rbrace $ denotes the set of solutions for $x$ satisfying $\gamma(x)=0$. Incorporating all these results into Eq.~$\eqref{eq:sigma12}$, we find
\begin{align}
{\mathcal S}_1 (\omega) & = {\mathcal C}_0 \, \sum_{n=J-1}^{\nmax} \, \sum_{k_0}
\Bigg[
\frac{f_{n,+}-f_{n+1,+}}{E_{n,+}(k_0)-E_{n+1,+}(k_0)}  \, 
	\frac{\Theta(\lambda_J^2 \, \dperm{n}{J} - \omega^2)}
	{\big| E_{n+1,+}^{\prime}(k_0)-E_{n,-}^{\prime}(k_0) \big|} \,
	\left (u_{n,+}^2 \, v_{n+1,+}^2 \, \perm{n}{J-1} \right ) 	
\nn & \hspace{ 2.75 cm} 	+  
 \frac{f_{n,+}-f_{n+1,-}}{E_{n,+}(k_0)-E_{n+1,-}(k_0)} \, 
\frac{\Theta(\omega^2 - \lambda_J^2 \, \dperm{n}{J})}
	{\big| E_{n+1,-}^{\prime}(k_0)+E_{n,+}^{\prime}(k_0) \big|} \,
\left(u_{n,+}^2 \, v_{n+1,-}^2 \, \perm{n+1}{J-1} \right) 
	\Bigg] \, ,
\nn 
{\mathcal S}_2 (\omega) & = {\mathcal C}_0 \, \sum_{n=J-1}^{\nmax} \, \sum_{k_0}
\Bigg[
\frac{f_{n+1,-}-f_{n,-}}{E_{n+1,-}(k_0)-E_{n,-}(k_0)}  \, 
	\frac{\Theta(\lambda_J^2 \, \dperm{n}{J} - \omega^2)}
	{\big| E_{n+1,-}^{\prime}(k_0)-E_{n,-}^{\prime}(k_0) \big|} \,
\left	(u_{n+1,-}^2 \, v_{n,-}^2 \, \perm{n}{J-1} \right ) 
\nn & \hspace{ 2.75 cm} 	+
\frac{f_{n+1,+}-f_{n,-}} {E_{n+1,+}(k_0)-E_{n,-}(k_0)}  \, 
	\frac{\Theta(\omega^2 - \lambda_J^2 \, \dperm{n}{J})}
	{\big| E_{n+1,+}^{\prime}(k_0) +E_{n,+}^{\prime} (k_0)\big|} \,
\left	(u_{n+1,+}^2 \, v_{n,-}^2 \, \perm{n}{J-1} \right ) \Bigg] \, ,
\end{align}
where $E^{\prime}(k_0) \equiv \partial E/ \partial k_z \vert_{k_z=k_0}$. The summation over $\lbrace k_0 \rbrace$ 
denotes the set of the roots of the equation
$$g(k_0) = \sqrt{
	\lsb \frac{\lambda_J^2 \left (\perm{n}{J} - \perm{\np}{J} \right ) + \omega^2}
	{2 \, \omega} \rsb^2 
- \lambda_J^2 \, \perm{n}{J}}\,, \text{ with } n'=n\pm 1, $$
which follows from Eq.~\eqref{eq:gkz}.

\bibliography{ref}

\end{document}